\documentclass[pdftex,twocolumn,epjc3]{svjour3}          

\RequirePackage[T1]{fontenc}

\smartqed  

\RequirePackage{graphicx}
\RequirePackage{mathptmx}      
\RequirePackage{flushend}
\RequirePackage[numbers,sort&compress]{natbib}
\RequirePackage[colorlinks,citecolor=blue,urlcolor=blue,linkcolor=blue]{hyperref}
\usepackage{subcaption}

\journalname{}

\begin{document}

\title{Properties of rotating neutron stars in light of binary compact object mergers}


\author{Bidisha Ghosh\thanksref{addr1}
        \and
        Mehedi Kalam\thanksref{e1,addr1} 
       }

\thankstext{e1}{e-mail: kalam@associates.iucaa.in}

\institute{Department of Physics, Aliah University, ll-A/27, Action Area II, Newtown, Kolkata -700160, India\label{addr1}
}

\date{Received: date / Accepted: date}

\maketitle

\begin{abstract}
The properties of rotating neutron stars are investigated using eight equations of state (EOSs). We also study the relations between various observables corresponding to different angular velocities for all those EOSs. 
All of these EoSs lead to non-rotating compact stars with maximum masses between 1.8 to 2.25 $M_{\odot}$. 
We calculate the moment of inertia and studied its variation with mass
and the relation between central energy density and angular momentum. 
We compare our results with the observational findings from the most 
massive pulsar PSR J0740+6620 and the heaviest secondary component in the
black hole - neutron star merger GW190814. It is noted that the secondary 
compact object of GW190814 having mass $\sim 2.6 M_{\odot}$ might be explained 
as a rapidly rotating neutron star (NS) with frequency larger than 1000 Hz. 
\keywords{Neutron star \and gravitational waves \and equation of states}
\end{abstract}

\section{Introduction}
On August 17, 2017 the LIGO and Virgo gravitational wave detectors, discovered the gravitational wave (GW) signal from the first binary neutron star (BNS) merger event GW170817. The detection of GW170817 provided 
the crucial information about the dense matter in neutron stars and its EoS.
From the gravitational wave data the chirp mass, 
$M_{chirp} = (m_1m_2)^{3/5}/(m_1+m_2)^{1/5}$, was accurately estimated to be 
1.188 $^{+0.04}_{-0.02} M_\odot$ \cite{r1}.  The masses of the binary component 
stars in GW170817 were in the range $1.17 - 1.60$ $M_{\odot}$ at 90$\%$ 
credible intervals and the total mass was 
$2.74_{-0.01}^{+0.04}$ $M_{\odot}$ for the low spin case that is consistent 
with the observed galactic BNSs. The GW signal from GW170817 had a duration of 
100 seconds and its characteristics in intensity and frequency created the 
expectation of the inspiral of two neutron stars. It was possible to estimate
the tidal deformability from the phase evolution of the GW signal during the 
inspiral. The post-merger signal could also reveal important aspects of EoSs and the fate of the compact remnant. But the LIGO and Virgo detectors are not sensitive to the post merger signal frequency that might be a few kHz. So, no post-merger signal was detected from GW170817\cite{r2}. The electromagnetic (EM) counterpart of GW170817 was also found. The kilonova associated with GW170817 was powered by the decay of radioactive nuclei synthesized in the ejected neutron-rich material. The fate can be explained by the amount of ejected material, estimated from the electromagnetic signals \cite{r3}. As the total mass of the system was 2.74 $M_\odot$ \cite{r1}, the remnant had undergone one of four possibilities- i) immediate
collapse to a black hole (BH), ii) a hypermassive neutron star (HMNS) supported by the differential rotation that collapsed to a BH on a timescale of a few tens of ms, iii) a supermassive neutron star (SMNS) and iv) a stable neutron star \cite{r4}-\cite{r5}. Last two cases involve comparatively less massive stars and could survive as uniformly rotating neutron stars. It was inferred 
that the remnant of GW170817 was short lived. The second BNS merger GW190425 had a chirp mass of $1.44 \pm 0.02 M_{\odot}$ \cite{r5b}. For the low spin case, the component masses at 90$\%$ credible intervals varied in the range 
1.46 - 1.87 $M_{\odot}$ whereas the range for the high spin case was 1.12 - 2.52 $M_{\odot}$.

On the other hand, the gravitational wave event GW190814 consisted of a black hole (BH) with a mass of $22.2-24.3$ $M_{\odot}$ and a compact object with a mass of $2.50-2.67$ $M_{\odot}$ at the borderline of heaviest neutron stars and
lightest black holes \cite{r6}. This secondary compact object could either be the least massive BH or the most massive NS. If the secondary becomes a NS, two possibilities might happen. According to Nathanail et al. \cite{r7}, this 
object would be a slowly rotating or non-rotating NS. But for this to happen, the maximum mass obtained by solving the Tolman-Oppenheimer-Volkoff (TOV) equation ($M_{TOV}$) should be $>$ 2.5 $M_{\odot}$.And the other possibility is a rapidly rotating NS whose maximum mass is enhanced by
$\sim 20\%$ above the maximum mass of a non-rotating star due to the rotation \cite{r8}-\cite{r9}. 
So, we investigate the case where the secondary component is a rapidly rotating NS.

Considering the isothermal temperature distribution over the most part of 
neutron stars and conserving the total baryonic mass, Hashimoto et al. 
\cite{r10} had shown that the maximum angular velocity could not exceed 
$7\times10^{3}$ $s^{-1}$. In papers \cite{r11} and \cite{r12}, the possibility 
of a hot and dense merger remnant was proposed from the binary neutron star 
system and it had also been numerically studied in \cite{r13}, \cite{r14},
and \cite{r15}. This 
motivates us to consider comparatively high angular velocity for the merger 
remnant in GW170817. Observational data provided sufficient evidence that 
the merger remnant did not collapse to a BH immediately, and it was a neutron 
star for a certain amount of time. So, analyzing the results of the rotation of 
the merger remnant we can have an idea about the secondary component of 
GW190814.

As a rotating object, one of its most important property is the 
moment of inertia (I). If the measurement of moment of inertia varies with 
$10\%$, the measured radius will vary within $6-7\%$ \cite{r16} for most of 
the cases. Thus one can accurately estimate the radius of a neutron star 
without any other astrophysical observation. That would put a strong constrain 
on the EoS of the neutron star. This makes us interested to study the 
behaviour of I with $M_\odot$ for different EoSs considering different angular 
velocities. For a given angular velocity, each equation of state will give a 
sequence of neutron stars upto the Keplerian limit. This Keplerian 
limit will allow us to correctly consider the rotational 
frequency to reach the proposed mass limit of the secondary compact object in 
GW190814. 
Here, in this paper, we discuss about the formalism and EoSs in section 2 and explain the results in section 3. Finally, section 4 contains discussion and 
conclusion.
  
\section{Formalism and Equation of States} 
A neutron star is a collapsed core of a very heavy mass star (>8$M_\odot$). It consists of five regions- atmosphere, envelope, crust, outer core and inner core. The atmosphere and envelope contain negligible amount of mass where the core region contains almost 99\% of the total mass. The core of a neutron star is still very mysterious to us. It has been considered that the outer core can be made of superfluid neutron and superconducting proton \cite{r17} and the inner core might have hyperons, Bose condensates (pions or kaons) and quarks play a major role over there. That is why Neutron stars are one of the most exotic objects in the universe. These stars are composed of so extremely condensed matter, that, to construct the relativistic models of these stars, the theory of general relativity should be used \citep{r18}, \cite{r19} and one needs to use the Einstein's field equations
\begin{eqnarray}
G^{{\mu}{\nu}} = R^{{\mu}{\nu}} - \frac{1}{2} g^{{\mu}{\nu}} R = 8 {\pi} T^{{\mu}{\nu}}~,
\end{eqnarray}
where the energy-momentum tensor 
\begin{eqnarray}
T^{{\mu}{\nu}}=({\epsilon}+P) u^{\mu} v^{\nu} + P g^{{\mu}{\nu}}~.
\end{eqnarray}
$G^{{\mu}{\nu}}$, $R^{{\mu}{\nu}}$, $g^{{\mu}{\nu}}$ and R are the Einstein curvature tensor, the Ricci tensor, the metric tensor and Ricci scalar respectively. P is the pressure and ${\epsilon}$ is the energy density, while $u^{\mu}$ is the matter four velocity. These field equations establish a strong connection between the general relativity and the strong interaction in dense matter \cite{r19}.
For a static star in the general relativity, under the condition of hydrostatic equilibrium, the above mentioned Einstein field equation reduces to the TOV equation \cite{r20} \cite{r21}
\begin{eqnarray}
\frac{dP (r)}{dr} = {-\frac{\epsilon(r) m (r)}{r^{2}}} {[1 + \frac{P(r)}{\epsilon (r)}]}
 {[1 + \frac{4 \pi r^{3} P (r)}{m (r)}]} \nonumber \\
 {[1 - \frac{2 m (r)}{r}]^{-1}}a~,
\end{eqnarray}
where $m(r)$ is the enclosed mass in a radius $r$, can be written as
\begin{eqnarray}
\frac{d m (r)}{d r} = {4} {\pi} {\epsilon (r)} {r^{2}}~,
\end{eqnarray}

The rapidly rotating neutron star (RNS) code \cite{r22} constructs models for rapidly rotating, relativistic neutron stars. So, for the case of a rapidly rotating neutron star the metric can be written as
\begin{eqnarray}
ds^{2} = -e^{\gamma + \rho}dt^{2} + e^{2\alpha} (dr^{2} + r^{2} d \theta^{2}) \nonumber \\
+e^{\gamma - \rho} r^{2} sin^{2}\theta(d\phi - \omega dt)^{2}~,
\end{eqnarray}
where the metric potential $\gamma$, $\rho$, $\alpha$, $\omega$ are functions of radial coordinate R and polar coordinate $\theta$ only \cite{r23}. This metric is considerably more complex than the case of a static one as the metric depends on the $\theta$ coordinate. And also, the rotation makes a neutron star stabilized against the gravitational collapse and therefore a rotating neutron star becomes more massive than a static one \cite{r24}. We consider that the matter source inside a rigidly rotating neutron star is a perfect fluid \cite{r25} and its energy-momentum tensor is given by equation (2). The proper velocity of matter with respect to local zero angular momentum observer (ZAMO) can be written as \cite{r26} 
\begin{eqnarray}
v\ = (\Omega - \omega) r sin {\theta} e^{- \rho}~,
\end{eqnarray}
where $\Omega = u^{3} / u^{0}$ is the angular velocity of the fluid. The four velocity of matter can be written as
\begin{eqnarray}
u^ {\mu} = \frac{e^ {- (\gamma + \rho) / 2}}{(1 - v\ ^{2})^{1 / 2}} (1, 0, 0, \Omega)~.
\end{eqnarray}

Substituting this equation into the Einstein field equations projected onto the local ZAMO frame, produces the three elliptical partial differential equations for metric potential $\rho$, $\omega$, $\gamma$ and the ordinary differential equation for the metric potential $\alpha$ \cite{r27} \cite{r28} \cite{r29}. By using Green's function method, the elliptical equations for the above three metric potentials are converted to integral equations.
From the relativistic equation of motion, the equation of hydrostatic equilibrium for a barotropic fluid may be obtained as \cite{r23}
\begin{eqnarray}
h (P) - h_{p} = \frac{1}{2} [{\gamma}_{p} + {\rho}_{p} - \gamma - \rho - {\ln} (1 - {v}^{2}) \nonumber \\
+ A^{2} (\omega - {\Omega}_{c})^{2}]~,
\end{eqnarray}
where $h(p)$ is the specific enthalpy, $h_{p}$ is the specific enthalpy at the pole, ${\gamma}_{p}$, ${\rho}_{p}$ are the values of the metric potential at pole, ${\Omega}_{c} = {\Omega} {r}_{e}$ and $A$ is the rotational constant \cite{r29}. The subscripts ${p}$, ${c}$ and ${e}$ denote the values at pole, center and equator respectively. The RNS code solves the ordinary differential equation for the metric potential ${\alpha}$ along with the converted integral equations for all the three metric potentials $\gamma$, $\omega$,$\rho$, equation (8) and the hydrostatic equilibrium equation at the pole [given by $h( {P}_c)$] and at the center [$h (P_e) = 0$] to calculate $\rho$, $\gamma$, $\omega$, $\alpha$, the angular velocity ($\Omega$), the equatorial coordinate radius ($r_{e}$), the density ($\epsilon$) and the pressure ($P$) profiles.\\                                    
 Here, we have computed different physical parameters using the RNS code. This code is written based on the Komatsu, Eriguchi and Hachisu method \cite{r30}. The RNS needs an EoS as the input the format of energy density (in $gm/cm^{3}$), pressure (in $dynes/cm^{2}$), enthalpy (in $cm^{2}$/$s^{2}$), baryon number density (in $cm^{-3}$) for its computation. Here, the field equations have been solved by fixing central energy density and any one of the following parameters, mass, rest mass, angular velocity, angular momentum, or the ratio of the polar coordinate radius to the coordinate equatorial radius.

We have studied four EoSs from the online data base in ref \citep{r31} using the RNS. The AP4 EoS is taken from \cite{r32},
ENG, SQM3 and ALF2 EoSs are from \cite{r33}, \cite{r34}, \cite{r35} respectively.\\
(1) The AP4 EoS is made of nucleons and obtained using the variational method where 2-body potential AV18, the UIX- 3 body potential and relativistic boost corrections are included \cite{r32}.\\
(2) The ALF2 is a hybrid EoS, consists of ap4 nuclear matter and a color-flavor-locked quark matter.\\
(3) The SQM3 EoS is represented by the strange quark matter based on the MIT bag model.\\
(4) The ENG EoS has been derived using the relativistic Dirac-Brueckner-Hartree-Fock approach.\\ 

The other set of EoSs used in this calculation consists of DD2, HS(TM1), SFHo and BHBLP, which have been studied using the numerical library LORENE \citep{r36}. Within this four EoSs, DD2, HS(TM1) and SFHo were developed by Hempel and Schaffner \cite{r37}, based on the statistical model along with the relativistic mean field (RMF) interactions. And the BHBLP EoS is basically the extended DD2 RMF interactions with $\Lambda$ hyperons interacting via $\phi$ messons \cite{r38}. The meson-baryon couplings in the DD2 and BHBLP EoSs are
density dependent. All of those are unified EoSs in the sense that the 
same RMF interactions are adopted below and above the saturation density.

\section{Results}
\begin{figure}
\includegraphics[width=0.5\textwidth]{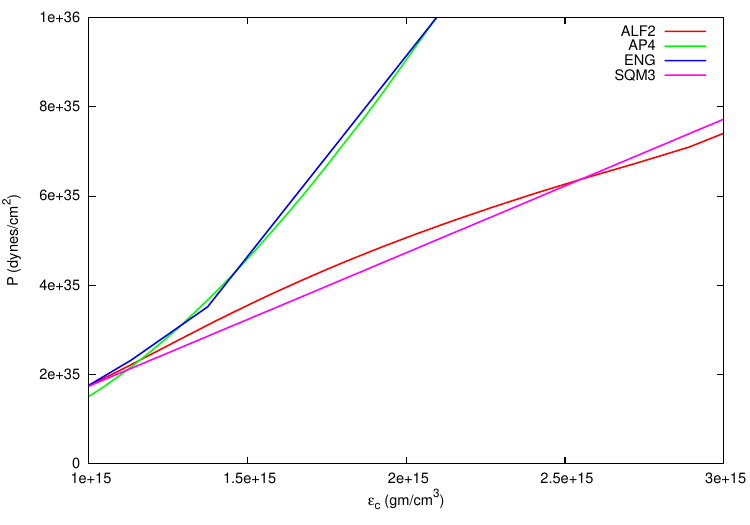}
\caption{Pressure versus energy density relation is shown for all the EoSs.}
\end{figure}
All our selected EoSs for the RNS code i.e. ALF2, AP4, ENG and SQM3 are shown in Fig. 1. The appearance of quarks in ALF2 and SQM3 makes those EoSs softer compared with two nuclear EoSs. Fig. 2 shows the neutron star mass as a function of radius for all the above mentioned EoSs where the maximum mass of the stars  varies in the range of 1.8 $\mathbf{ M_\odot}$ - 2.25 $\mathbf{M_\odot}$. Observations of galactic pulsars led to the discoveries of massive neutron stars such as PSR J0740+6620 of 2.08 $\pm$ 0.07 $M_{\odot}$ and PSR J0348+0432 with 2.01 $\pm$ 0.04 $M_{\odot}$  \cite{r39}, \cite{r40}, \cite{r41}. This gives the lower bound on the maximum mass of neutron stars. Recently x-ray observational data from the Neutron Star Interior Composition Explorer (NICER) and X-ray Multi-Mirror (XMM-Newton) were also used to estimate the radius and the mass of PSR J0740+6620 which is in agreement with the value of the radio observation \cite{r42}, \cite{r43}, \cite{r44}. 

For the merger remnant of GW170817, Shunke et al.  \cite{r45} theoretically reached to this conclusion that if the merger product is a HMNS, then $M_{TOV} < 2.09^{+0.06}_{-0.04}(^{+0.11}_{-0.09})$ $M_{\odot}$; if the merger product is a SMNS, then the final result will be within the limit of $2.09^{+0.06}_{-0.04}(^{+0.11}_{-0.09})$ $M_{\odot} \leq M_{TOV} < 2.43^{+0.06}_{-0.04}(^{+0.10}_{-0.08})$ $M_{\odot}$ at the confidence level of 2$\sigma$ (1$\sigma$). And for the case of a SNS, $M_{TOV} \geq 2.43^{+0.06}_{-0.04}(^{+0.10}_{-0.08})$ $M_{\odot}$ will be the limit. 
\begin{figure}
\includegraphics[width=0.5\textwidth]{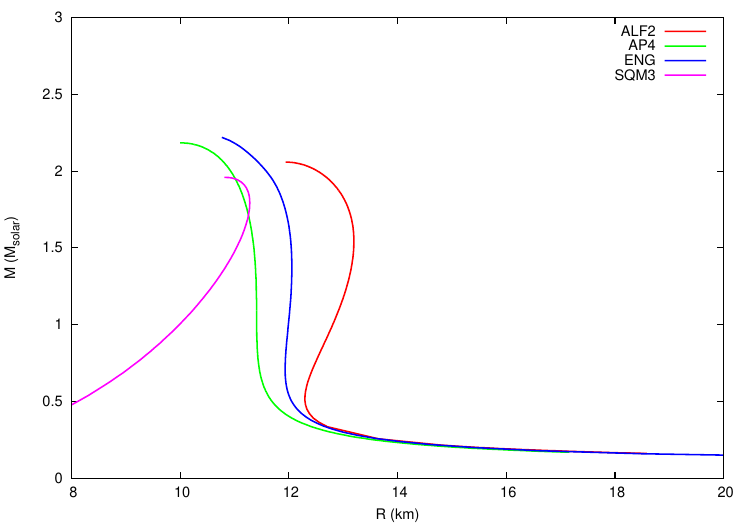}
\caption{Mass versus radius relation is shown for all the EoSs.}
\end{figure}

We have investigated the properties of rapidly rotating neutron stars using the RNS code. Here we show our results for three different angular velocities - $4\times10^{3} s^{-1}$, $5\times10^{3} s^{-1}$ and $6\times10^{3} s^{-1}$ within the central energy density range $1\times10^{15}$ to $3\times10^{15}$ $gm/cm^{3}$. The selected angular velocities do not exceed the maximum angular velocity range that was predicted in \cite{r10}. In Fig. 3[(a), (b) and (c)], the variation of the moment of inertia with neutron star mass is shown for angular velocity - $4\times10^{3} s^{-1}$, $5\times10^{3} s^{-1}$ and $6\times10^{3} s^{-1}$ respectively. From the value of the moment of inertia and mass, it is possible to estimate the value of the radius, thus we can put a constrain on EoS. So, the moment of inertia plays a very important role in determining the EoS.                             

The values of the moment of inertia, mass and radius and compactness are shown in tables 1-3 for our selected angular velocities. For a particular EoS, the $I$ is higher for a star rotating at angular velocity $6 \times 10^{3} s^{-1}$ than a star, rotating at $4 \times 10^{3} s^{-1}$. 
Also the data in the tables show that the value of $I$ is higher for the lower energy density value and lower for the higher energy density value for the ALF2 and SQM3 EoSs which are softer than the other cases. Although the EoS ENG shows this type of behaviour for angular velocity $6 \times 10^{3} s^{-1}$, but the opposite is true for other angular velocities.\\

On the other hand, the compactness ($M/R$) parameter clearly depends on EoSs. It varies differently for the selected EoSs. According to Buchdahl \cite{r46}, maximum allowable mass-radius ratio for a spherically symmetric perfect fluid sphere should be $\frac{Mass}{Radius} < \frac{4}{9}$. We see from the tables that it does not depend on the angular velocity; at any particular energy density it shows the same value for all the velocities. 

\begin{table}[th!]
\caption{EoS properties for $\Omega = 4\times10^{3} s^{-1}$}
\begin{center}
\begin{tabular}{c c c c c c c}
\hline\hline\vspace{-0.4cm}\\
EoS  & $M_{\odot}^{\epsilon_{1}}$ & $M_{\odot}^{\epsilon_{2}}$ & $I_{\epsilon_{1}}$ & $I_{\epsilon_{2}}$ & $C_{\epsilon_{1}}$ & $\mathbf{C_{\epsilon_{2}}}$\\
\hline
ALF2 & 1.98              & 2.10              & 2.90     & 2.41   & 0.15  & 0.19\\
AP4  & 1.49              & 2.24              & 1.54     & 2.26   & 0.13  & 0.23\\
ENG    & 1.75            & 2.27              & 2.10     & 2.39   & 0.14  & 0.22\\
SQM3  & 1.85             & 1.99              & 2.41     & 2.09   & 0.15  & 0.18\\

\end{tabular}

\end{center}
Here $\epsilon_{1} = 1\times10^{15} gm/cm^{3}$ and ${\epsilon}_{2} = 3 {\times} 10^{15} gm/cm^{3}$, I is in $10^{45} gm/cm^{3}$ and C ($M_{\odot}/ R$) is called compactness
\end{table}

\begin{figure*}[ht]
\begin{center}
\begin{subfigure}{0.33\textwidth}
\includegraphics[width=\textwidth]{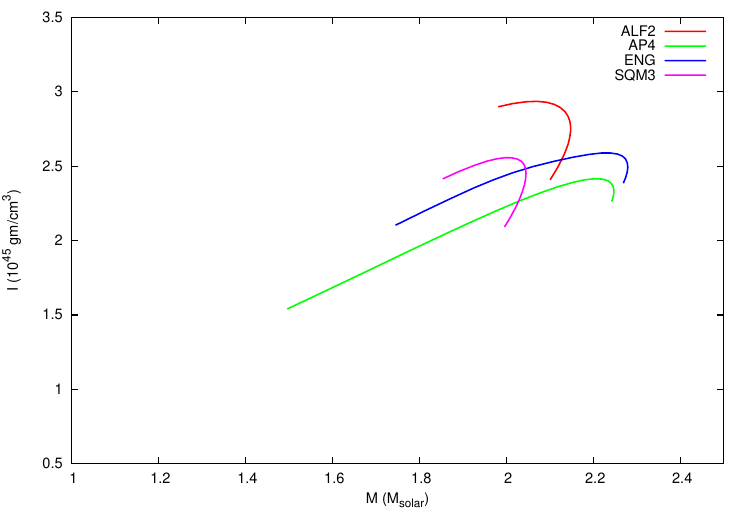}
\caption{}
\label{fig:subim 1}
\end{subfigure}
\begin{subfigure}{0.33\textwidth}
\includegraphics[width=\textwidth]{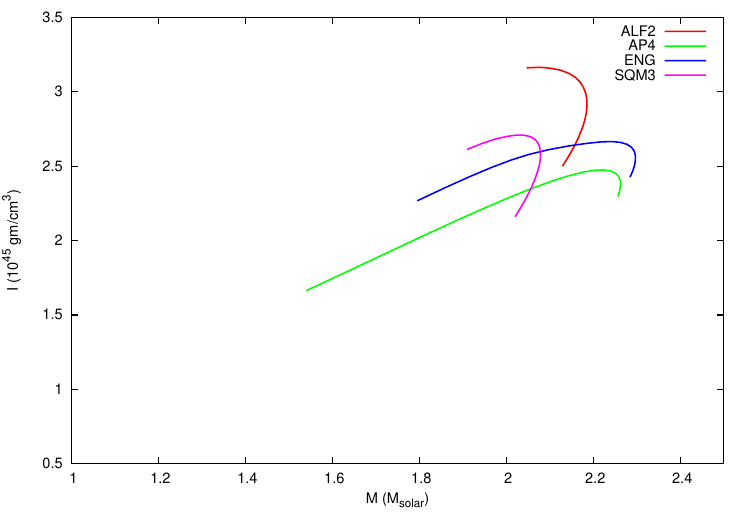}
\caption{}
\label{fig:subim 2}
\end{subfigure}
\begin{subfigure}{0.33\textwidth}
\includegraphics[width=\textwidth]{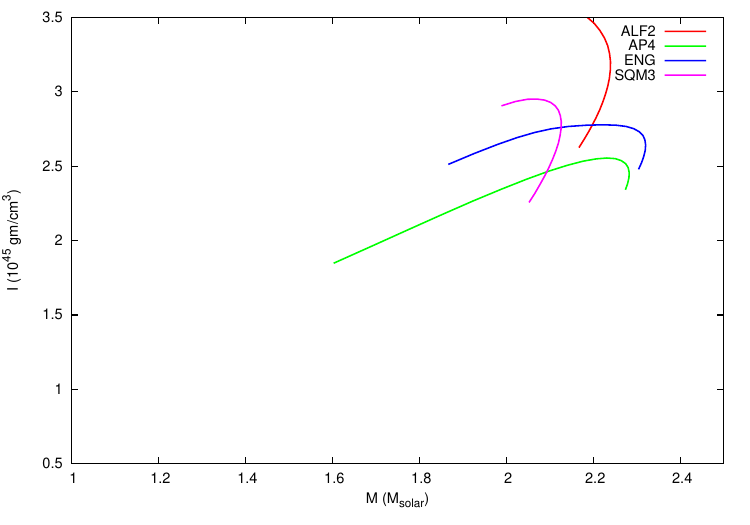}
\caption{}
\label{fig:subim 3}
\end{subfigure}
\caption{Moment of inertia versus mass relation is shown for all the EoSs for angular velocity $4\times10^{3} s^{-1}$, $5\times10^{3} s^{-1}$,  $6\times10^{3} s^{-1}$ in figures (a), (b), (c) respectively.}.
\end{center}
\end{figure*}

\begin{table}[th!]
\caption{EoS properties for $\Omega = 5\times10^{3} s^{-1}$}
\begin{center}
\begin{tabular}{c c c c c c c}
\hline\hline\vspace{-0.4cm}\\
EoS  & $M_{\odot}^{\epsilon_{1}}$ & $M_{\odot}^{\epsilon_{2}}$ & $I_{\epsilon_{1}}$ & $I_{\epsilon_{2}}$ & $C_{\epsilon_{1}}$ & $\mathbf{C_{\epsilon_{2}}}$\\
\hline
ALF2 & 2.05              & 2.13              & 3.16     & 2.50   & 0.15  & 0.18\\
AP4  & 1.54              & 2.26              & 1.66     & 2.29   & 0.13  & 0.23\\
ENG  & 1.79              & 2.28              & 2.26     & 2.42   & 0.14  & 0.22\\
SQM3 & 1.91              & 2.02              & 2.61     & 2.16   & 0.15  & 0.18\\

\end{tabular}

\end{center}
Here ${\epsilon}_{1} = 1 {\times} 10^{15} gm/cm^{3}$ and ${\epsilon}_{2} = 3 {\times} 10^{15} gm/cm^{3}$, I is in $10^{45} gm/cm^{3}$ and C ($M_{\odot}/ R$) is called compactness
\end{table}
\begin{table}[th!]
\caption{EoS properties for ${\Omega} = 6 {\times} 10^{3} s^{-1}$} 
\begin{center}
\begin{tabular}{c c c c c c c}
\hline\hline\vspace{-0.4cm}\\
EoS  & $M_{\odot}^{\epsilon_{1}}$ & $M_{\odot}^{\epsilon_{2}}$ & $I_{\epsilon_{1}}$ & $I_{\epsilon_{2}}$ & $C_{\epsilon_{1}}$ & $\mathbf{C_{\epsilon_{2}}}$\\
\hline
ALF2 & 2.14              & 2.17              & 3.57     & 2.63   & 0.15  & 0.18\\
AP4  & 1.60              & 2.27              & 1.85     & 2.34   & 0.13  & 0.23\\
ENG  & 1.87              & 2.30              & 2.51     & 2.48   & 0.14  & 0.22\\
SQM3 & 1.98              & 2.05              & 2.90     & 2.26   & 0.15  & 0.18\\

\end{tabular}

\end{center}
Here ${\epsilon}_{1} = 1 {\times}10^{15} gm/cm^{3}$ and ${\epsilon}_{2} = 3 {\times}10^{15} gm/cm^{3}$, I is in $10^{45} gm/cm^{3}$ and C ($M_{\odot}/ R$) is called compactness
\end{table}
Fig. 4[(a), (b), (c)] show the variation of angular momentum with central energy density for all the EoSs. At energy density $1 {\times} 10^{15} gm/cm^{3}$, the value of angular momentum is higher for ALF2, SQM3 EoSs for all the selected angular velocities. It is also very interesting to see that for the other EoSs, the angular momentum is lower at $1 \times 10^{15} gm/cm^{3}$ but slightly increases after that and becomes almost flat near $2 \times 10^{15} gm/cm^{3}$. But for ALF2 and SQM3 EoSs, it shows higher value at lower energy density and starts decreasing approximately after $\mathbf{1.5 \times 10^{15} gm/cm^{3}}$ except for EoS ALF2 at angular velocity $\mathbf{6 \times 10^{3} s^{-1}}$, where it starts decreasing from the very initial value.

\begin{figure*}[th]
\centering
\begin{subfigure}{0.33\textwidth}
\includegraphics[width=\textwidth]{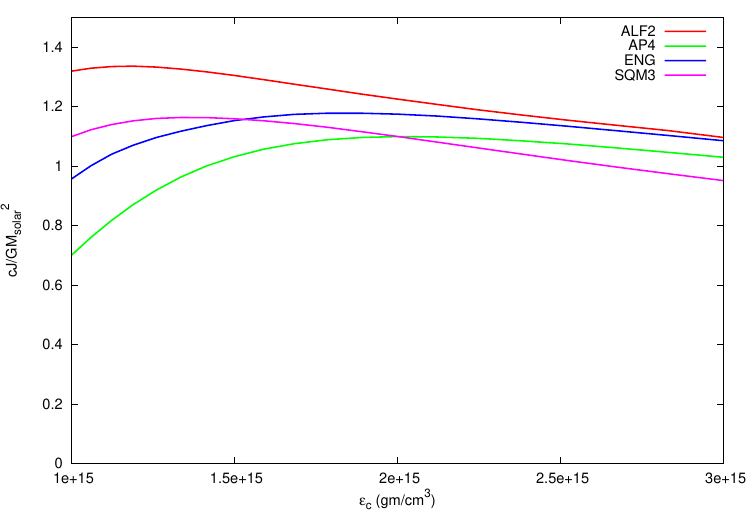}
\caption{}
\label{fig:subim 1}
\end{subfigure}
\begin{subfigure}{0.33\textwidth}
\includegraphics[width=\textwidth]{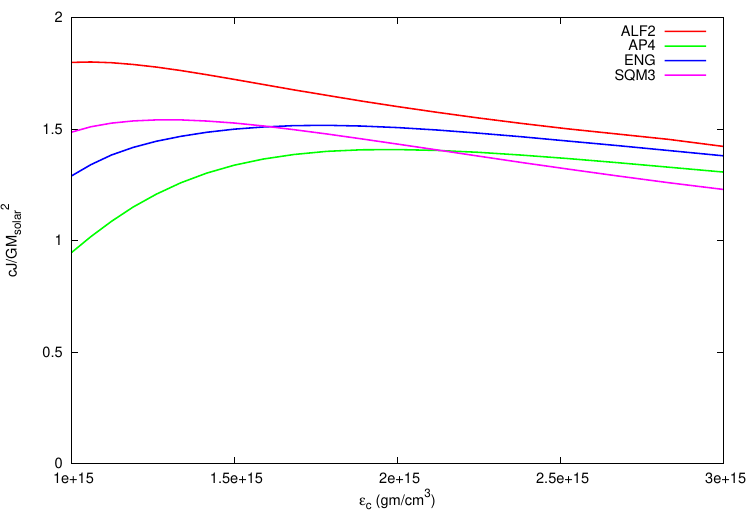}
\caption{}
\label{fig:subim 2}
\end{subfigure}
\begin{subfigure}{0.33\textwidth}
\includegraphics[width=\textwidth]{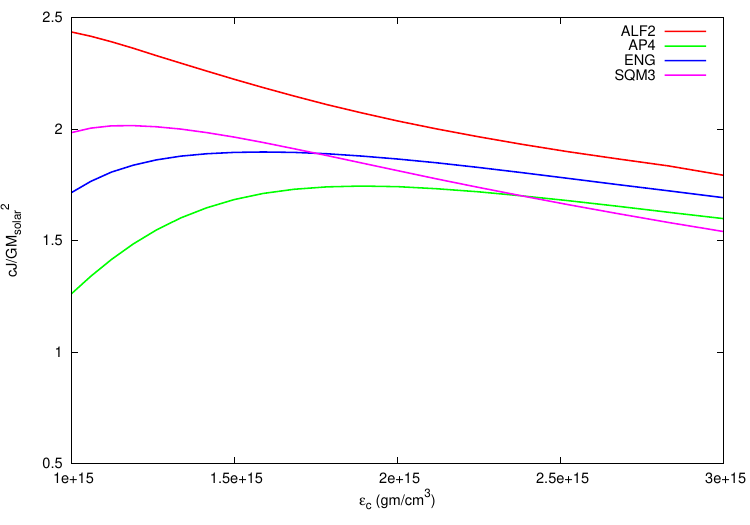}
\caption{}
\label{fig:subim 3}
\end{subfigure}
\caption{Angular momentum versus central energy density is shown for all the EoSs for angular velocity $4\times10^{3} s^{-1}$, $5\times10^{3} s^{-1}$, $6\times10^{3} s^{-1}$ in figures (a), (b), (c) respectively.}
\end{figure*}
Assuming the merger remnant was short lived and collapsed to a BH finally, different groups estimated the mass of the merger remnant at the time of the collapse in the range 2.17 - 2.3 $M_{\odot}$ \cite{r40}, \cite{r41}, \cite{r47}, \cite{r48}. This sets the upper bound on the maximum mass of neutron stars. All EoSs studied here are compatible with the upper bound of the maximum mass as found by Ruiz et al. \cite{r40}, Margalit and Metzger \cite{r41}, Rezzolla et al. \cite{r47}, and Shibata et al. \cite{r48}. Table 1, 2 and 3 have shown us how the mass of a particular star corresponding to an EoS evolves with the variation of the angular velocity. We further investigate the uniformly rotating compact stars using the other set of EoSs in the following paragraph.

Using the numerical library Lorene, we have studied the rotating stars with the DD2, SFHo, HS(TM1) and BHBLP EoSs of the beta-equilibrated and charge neutral matter. In table 4, $M_{TOV}^{max}$ masses are shown, where all EoSs except the DD2 EoS are compatible with the $M_{TOV}^{max}$ limit set by GW170817 \cite{r40}, \cite{r41}, \cite{r47}, \cite{r48}. The second line of table 4 shows the maximum mass at the Keplerian frequency. Taking into consideration the fact that the rotation increases the mass of a star by 20$\%$ \cite{r8}, \cite{r9}, we have computed these with the beta equilibrated EoSs, and showed our results in Fig. 5[(a),(b),(c) and (d)]. The maximum mass for all of these EoSs at the Keplerian frequency is varying within the range of $\sim 2.45-2.92 M_{\odot}$. 

The secondary compact object of GW190814 whose mass ranges between 2.50-2.67 $M_{\odot}$ \cite{r6}, could be explained as a uniformly rotating massive neutron star if the rotational frequency is greater than 1000 Hz. In Fig. [6 and 7], the variation of mass and radius can be seen for the Keplerian limit and the non-rotating case, respectively. According to our analysis, the radius of the compact star will be within the limit of 10.98 km to 14.63 km if the mass ranges between 2.50 - 2.67 $M_{\odot}$. These values of the radius are obtained for all EoSs except the SFHo EoS. The maximum mass for the SFHo EoS is $\sim 2.44 M_{\odot}$ and the radius is 9.53 km at the Keplerian limit. 
\begin{table}[th!]
\caption{The maximum mass ($M^{max}$) values for the non-rotating case and at the Keplerian frequency respectively.}
\begin{center}
\begin{tabular}{c c c c c}
\hline\hline\vspace{-0.4cm}\\
EoS  & DD2 & HS(TM1) & SFHo & BHBLP\\
\hline
$M_{TOV}^{max}$ & 2.42206 $M_{\odot}$ & 2.21279 $M_{\odot}$ & 2.05817 $M_{\odot}$ & 2.10024 $M_{\odot}$\\
$M_{Kep}^{max}$ & 2.91587 $M_{\odot}$ & 2.64659 $M_{\odot}$ & 2.44320 $M_{\odot}$ & 2.52157 $M_{\odot}$\\ 

\end{tabular}
\end{center}
\end{table}

\begin{figure*}[th]
\centering
\begin{subfigure}{0.33\textwidth}
\includegraphics[width=\textwidth]{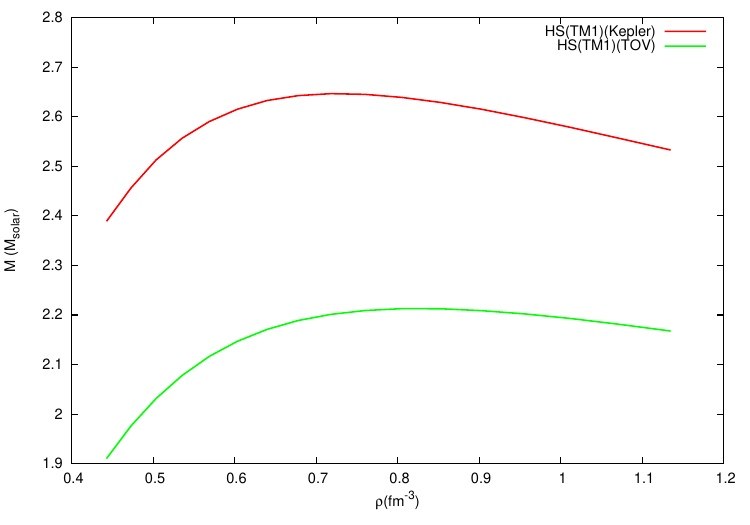}
\caption{}
\label{fig:subim 1}
\end{subfigure}
\begin{subfigure}{0.33\textwidth}
\includegraphics[width=\textwidth]{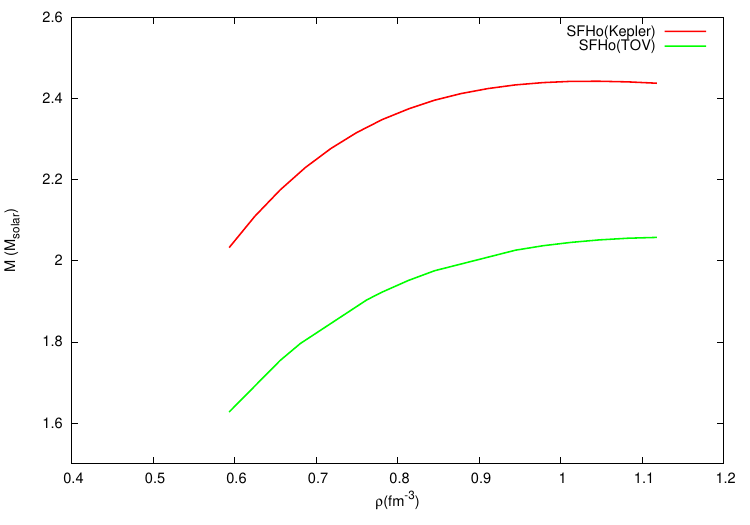}
\caption{}
\label{fig:subim 2}
\end{subfigure}
\begin{subfigure}{0.33\textwidth}
\includegraphics[width=\textwidth]{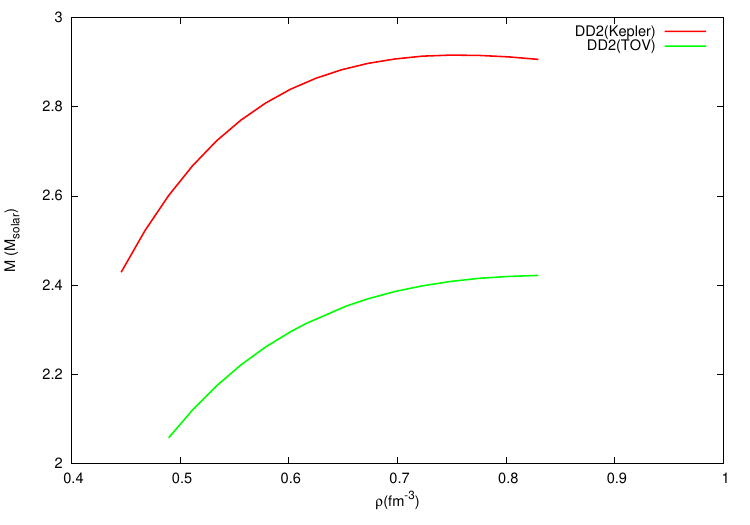}
\caption{}
\label{fig:subim 3}
\end{subfigure}
\begin{subfigure}{0.33\textwidth}
\includegraphics[width=\textwidth]{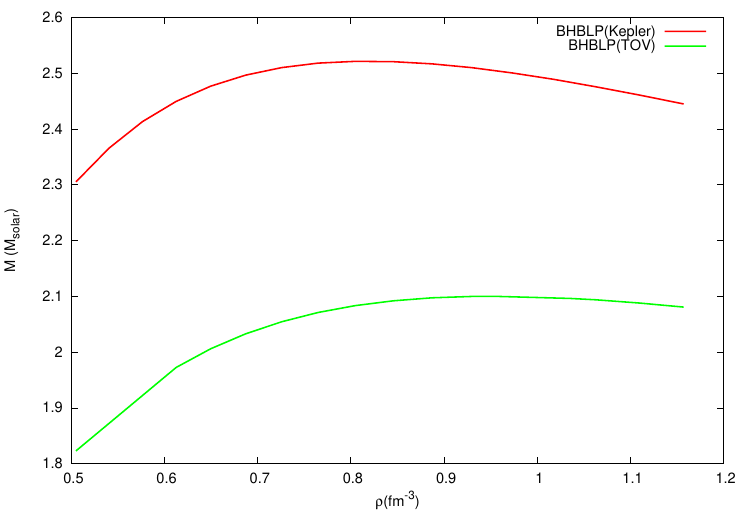}
\caption{}
\label{fig:subim 4}
\end{subfigure}
\caption{Mass versus baryon density for the EoSs HS(TM1), SFHo, DD2, BHBLP is shown in figures (a), (b), (c), (d) respectively. The red line indicates the sequence of rotating neutron stars at the Keplerian frequency. And the green line is for the non-rotating case.}
\end{figure*}

\begin{figure}
\includegraphics[width=0.4\textwidth]{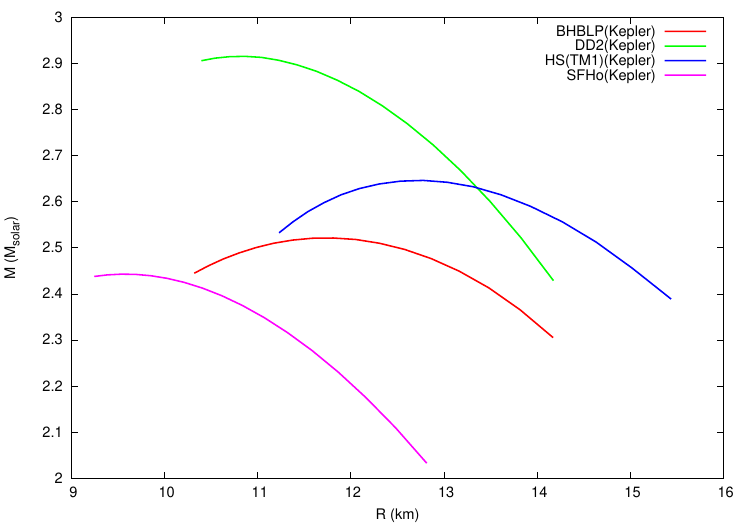}
\caption{Mass versus radius at the Keplerian frequency}
\end{figure}
\begin{figure}
\includegraphics[width=0.4\textwidth]{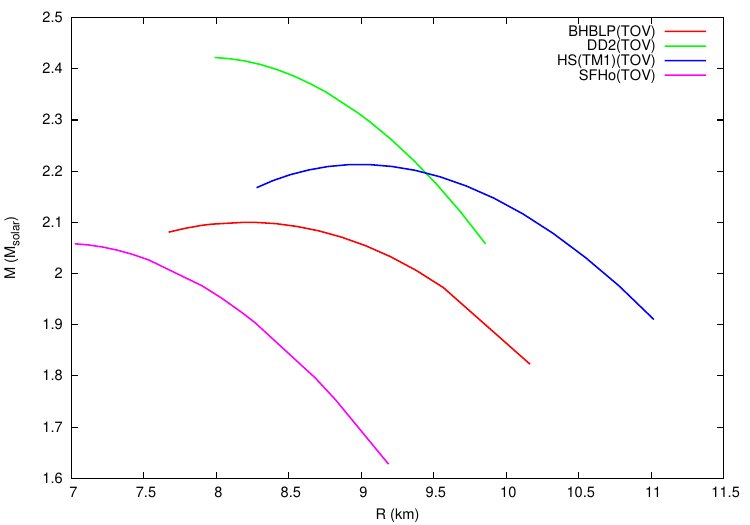}
\caption{Mass versus radius for the non-rotating case}
\end{figure}
According to the observation, the most massive neutron star till date, PSR J0740+6620, a rotation powered millisecond pulsar with a binary companion, rotates with a frequency at 346 Hz approximately. Miller et al. analyzed the recent observational data of the NICER and XMM and estimated its mass and radius to be $2.08 {\pm} 0.07$ $M_{\odot}$ \cite{r42} and $13.7_{-1.5}^{+2.6}$ km respectively \cite{r43}, but Riley et al. in \cite{r49} constrained the mass and radius of PSR J0740+6620 to be $2.072_{-0.066}^{+0.067}$ $M_{\odot}$ and $12.39_{-0.98}^{+1.30}$ km respectively. So, we have computed the mass-radius relation using our EoSs with this observed frequency and the result is shown in Fig. 8. For a star of mass around 2.1 $M_{\odot}$, we get the radius varying between 9.36 km to 11.91 km except the SFHo EoS, which is almost similar to the value as stated in \cite{r49}. 
\begin{figure}
\includegraphics[width=0.4\textwidth]{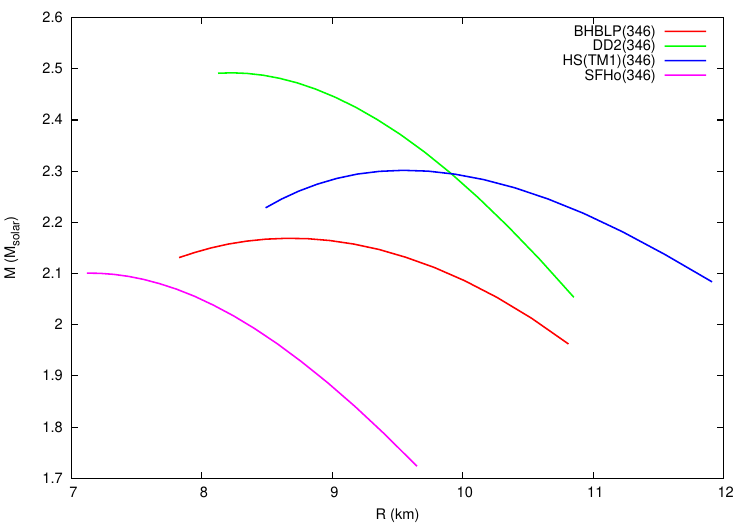}
\caption{Mass versus Radius for all the EoSs at frequency 346 Hz}
\end{figure}

\section{Discussion and Conclusion}
 We have computed here the mass-radius relation, moment of inertia, compactness, angular momentum parameters of uniformly rotating neutron stars considering eight equations of state. 
Among the AP4, ENG, ALF2 and SQM3 EoSs, we find that the AP4 and ENG EoSs are stiffer while  other two EoSs are softer due to the presence of quarks. 
 Our results have covered a sufficiently large range of the TOV maximum mass for all the cases to vary within the range of 1.8 $\mathbf{ M_{\odot}}$ to 2.25 $\mathbf{ M_{\odot}}$. One of the most important property for a rotating object is the moment of inertia which is EoS dependent. We have computed here the values of $I$ for all the EoSs for our three selected angular velocities. We note that the $I$ increases with increasing angular velocity for all our EoSs. 
Also we have shown here the variation of angular momentum with the central energy density. We find that, at the very initial central energy density value, angular momentum is lower for EoSs AP4 and ENG than the other two EoSs for all angular velocities. But towards the ending, EoS SQM3 has the lowest angular momentum value for all the cases. 

Our results for the AP4, ENG, ALF2, SQM3, EoSs are consistent with the upper limit for the maximum $M_{TOV}$ mass of the merger remnant in GW170817 predicted by Ruiz et al. \cite{r40}, Margalit and Metzger \cite{r41}, Rezzolla et al. \cite{r47}, and Shibata et al. \cite{r48}. The DD2, HS(TM1), SFHo and BHBLP EoSs have been selected for the next part of our analysis, where we want to know about the nature of the secondary compact object of GW190814. It followed from the observation that the mass of the secondary component might be 2.50-2.67 $M_{\odot}$ \cite{r6}. The maximum possible gravitational mass for a non-rotating star is $M_{TOV}$ ${\simeq}$ 2.3 $M_{\odot}$ which is the upper limit set by GW170817. Considering the fact that the secondary object is a rapidly rotating NS, we have computed the maximum masses at the Keplerian limit for the DD2, SFHo, HS(TM1) and BHBLP EoSs. It is possible to reach the limit of 2.5-2.67 $M_{\odot}$ when the rotational frequency would be more than 1000 Hz for all the EoSs. And the corresponding radii have values 10.98-14.63 km. \\
PSR J0740+6620 is the most massive pulsar detected till date, with a mass of $2.08 {\pm} 0.07$ $M_{\odot}$ 
This pulsar rotates with a frequency of 346 Hz. When we compute the mass-radius relations for our EoSs with this frequency, we obtain its radius varying between 9.36-11.91 km for that given mass value of the pulsar. This result is consistent with observations. 
\section{Acknowledgement}
BG would like to express sincere gratitude to her advisor Prof. Debades Bandyopadhyay for his continuous support, valuable discussions, suggestions and corrections. BG thanks Dr. Debarati Chatterjee for helping with the programming codes and grateful to IUCAA, Pune, India, where a part of this work has been done. BG also thanks Aliah University for financial support.MK like to thank IUCAA, Pune, India
for providing research facilities under Visiting Associateship programme.


\begin{thebibliography}{9}

\bibitem{r1}
B. P. Abbott et al, PRL 119, 161101 (2017)
\bibitem{r2}
B. P. Abbott et al, Physical Review X 9, 011001, (2019)
\bibitem{r3}
N. Bucciantini, B. D. Metzger, T. A. Thompson, E. Quataert, MNRAS, 419, 1537, 2012
\bibitem{r4}
B. Giacomazzo, R. Perna, ApJL, 771, L26 (2013)
\bibitem{r5}
B. Margalit and B. D. Metzger, Astrophys. J. Lett. 850, L19 (2017).
\bibitem{r5b}
B. P. Abbott et al, Astrophys. J. Lett., {\bf 892} (2020) L3
\bibitem{r6}
R. Abbottet, et al., The Astrophysical Journal, 896, 2006.12611,  L44 (2020) 
\bibitem{r7}
A. Nathanail, E. R. Most, L. Rezzolla, The
Astrophysical Journal, 908, L28 (2021).
\bibitem{r8}
C. Breu, L. Rezzolla, MNRAS, 459, 646 (2016)
\bibitem{r9}
D.-S. Shao, S.-P. Tang, X. Sheng, et al., PhRvD, 101, 063029 (2020)
\bibitem{r10}
M. A. Hashimoto, K. Oyamatsu, Y. Eriguchi, The Astrophysical Journal, 436:257-261, (1994)
\bibitem{r11}
A. Endrizzi et al., arXiv:1908.04952
\bibitem{r12}
Y. Sekiguchi, K. Kiuchi, K. Kyutoku, M. Shibata, Phys. Rev. Lett., 107, 051102 (2011)
\bibitem{r13}
R. Cioffi, W. Kastaun, J. V. Kalinani, B. Giacomazzo, Phys. Rev. D, 100, 023005 (2019)
\bibitem{r14}
M. Shibata, et al., Phys. Rev. D, 96, 123012 (2017)
\bibitem{r15}
D. Radice, S. Bernuzzi, W. Del Pozzo, L. F Roberts, C. D. Ott, Astrophys. J.,
842, L10 (2017)
\bibitem{r16}
J. M. Lattimer and B. F. Schutz, Astrophys. J. 629 (2005) 979.
\bibitem{r17}
J. M. Lattimer, M. Prakash, Science, Vol. 304, Issue 5670, pp. 536-542, 2004
\bibitem{r18}
N. K. Glendenning, Compact Stars, Nuclear Physics, Particle Physics, and General Relativity (New York: Springer-Verlag), 2000.
\bibitem{r19}
F. Weber, Pulsars as Astrophysical Laboratories for Nuclear and Particle Physics (Bristol, Great Britan: IOP Publishing), 1999.
\bibitem{r20}
R. C. Tolman, Proc. Natl. Acad. Sci. U.S.A. 20, 3 (1934).
\bibitem{r21}
J. R. Oppenheimer, G. M. Volkov, Phys. Rev. 55, 374 (1939).
\bibitem{r22}
N. Stergioulas and J. L. Friedman, Astrophys. J. 444 (1995) 306.
\bibitem{r23}
G. B. Cook, S. L. Shapiro, S. A. Teukolsky, Astrophysical Journal, Part 1 (ISSN 0004-637X), vol. 424, no. 2, p. 823-845
\bibitem{r24}
P. G. Krastev, B. Li, A. Worley, The Astrophysical Journal, 676, 1170-1177, 2008
\bibitem{r25}
N. Stergioulas, 1996, Ph.D. thesis, Univ. Wisconsin, Milwaukee, 2003, Living Rev. Relativ., 6,
\bibitem{r26}
R. Ouyed, Astronomy and Astrophysics 382, 939-946 (2002) 
\bibitem{r27}
J. M. Bardeen, R. V. Wagoner, 1971, ApJ, 167, 359.
\bibitem{r28}
E. M. Butterworth, J. R. Ipser, 1976, ApJ, 204, 200 
\bibitem{r29}
H. Komatsu, Y. Eriguchi, I. Hachisu, 1989, MNRAS, 237, 355 
\bibitem{r30}
H. Komatsu, Y. Eriguchi and I. Hachisu, Mon. Not. R. Astron. Soc. 237 (1989) 355.
\bibitem{r31}

{\url{http://xtreme.as.arizona.edu/neutronstars/}}
\bibitem{r32}
A. Akmal, Pandharipande, V.R. 1997, Phys. Rev. C, 56, 2261
\bibitem{r33}
Engvik et al., Astrophysical Journal v.469, p.794, 1996
\bibitem{r34}
M. Prakash, J. R. Cooke, and J. M. Lattimer, Phys. Rev. D, 52, 661, 1995
\bibitem{r35}
M. Alford, M. Brady, M. Paris, S. Reddy, Astrophys Journal, 629, 969, (2005)
\bibitem{r36}
{\url{http://www.lorene.obspm.fr/}}
\bibitem{r37}
M. Hempel and J. Schaffner-Bielich, Nucl. Phys. A 837, 210 (2010).
\bibitem{r38}
S. Banik, M. Hempel and D. Bandyopadhyay, ApJS 214, 22 (2014).
\bibitem{r39}
L. Rezzolla, E. R. Moist and L. R. Weith, Astrophys. J. Lett. 852, L25 (2018)
\bibitem{r40}
M. Ruiz, S. L. Shapiro and A. Tsokaros, Phys. Rev. D 97, 021501 (2018)
\bibitem{r41}
B. Margalit and B. D. Metzger, Astrophys. J. Lett. 850, L19 (2017)
\bibitem{r42}
E. Fonseca, H. T. Cromartie, T. T. Pennucci, et al., arXiv:2104.00880, 2021
\bibitem{r43}
M. C. Miller, et al., 	arXiv:2105.06979
\bibitem{r44}
G. Raaijmakers, et al., arXiv:2105.06981
\bibitem{r45}
Shunke Ai, He Gao, Bing Zhang, The Astrophysical Journal, 893:146, 2020
\bibitem{r46}
H.A. Buchdahl, Phys. Rev. 116, 1027 (1959)
\bibitem{r47}
L. Rezzolla, E. R. Most, L. R. Weih, ApJL, 852, L25, 2018
\bibitem{r48}
M. Shibata, E. Zhou, K. Kiuchi, S. Fujibayashi, PhRvD, 100, 023015, 2019
\bibitem{r49}
T. E. Riley, A. Watts, P. Ray, et al., ApJL submitted, 2021
\bibitem{r50}
H. T. Cromartie, E. Fonseca, S. M. Ransom, et al., Nature Astronomy, 4, 72, 2020
\end{thebibliography}
\end{document}